\documentclass[aps,10pt,prr,twocolumn,superscriptaddress,floatfix,nofootinbib,longbibliography]{revtex4-2}

\usepackage{graphicx,amsfonts,amssymb,amsmath,hyperref}
\usepackage[ruled,vlined]{algorithm2e}
\usepackage{tikz}
\usetikzlibrary{positioning,arrows.meta}
\usepackage{lmodern}
\usepackage{enumitem}
\usepackage{multirow}
\usepackage{booktabs}
\usepackage[caption=false]{subfig}
\usepackage{ragged2e}
\usepackage{etoolbox}
\usepackage{dsfont}
\usepackage{array}
\usepackage[dvipsnames]{xcolor}
\usepackage[normalem]{ulem}

\graphicspath{{images/}}

\newif\ifhyper
\hypertrue
\ifhyper
\hypersetup{
   citecolor = {blue},
   colorlinks = {true},
   urlcolor = {blue}
}
\fi

\newcommand{\beq}{\begin{equation}}
\newcommand{\eeq}{\end{equation}}

\begin{document}

\title{Pareto-optimal quantum kernel selection for unsupervised anomaly detection on real malware beaconing data}

\author{Boaz Micah}
\affiliation{Multiverse Computing, Parque Científico y Tecnológico de Gipuzkoa, Paseo de Miramón 170, Planta 2, 20014 Donostia / San Sebastián, Spain}

\author{Nadia Milazzo}
\email{nadia.milazzo@iqm.tech}
\affiliation{IQM Quantum Computers, 4 rue Royale, 75008 Paris, France}

\author{Maissa Beji}
\affiliation{Multiverse Computing, 7 rue de la Croix Martre, 91120 Palaiseau, Paris, France}

\author{Borja Aizpurua}
\affiliation{Multiverse Computing, Parque Científico y Tecnológico de Gipuzkoa, Paseo de Miramón 170, Planta 2, 20014 Donostia / San Sebastián, Spain}
\affiliation{Department of Basic Sciences, Tecnun -- University of Navarra, San Sebasti\'{a}n, Spain}

\author{Lloren\c{c} Espinosa-Portal\'{e}s}
\email{llorenc.espinosa@multiversecomputing.com}
\affiliation{Multiverse Computing, Parque Científico y Tecnológico de Gipuzkoa, Paseo de Miramón 170, Planta 2, 20014 Donostia / San Sebastián, Spain}

\author{Esteban Payares}
\affiliation{IQM Quantum Computers, 4 rue Royale, 75008 Paris, France}

\author{Ghada Ben Slama}
\affiliation{IQM Quantum Computers, 4 rue Royale, 75008 Paris, France}

\author{Luc Andrea}
\affiliation{Multiverse Computing, 7 rue de la Croix Martre, 91120 Palaiseau, Paris, France}

\author{Michel Kurek}
\affiliation{Multiverse Computing, 7 rue de la Croix Martre, 91120 Palaiseau, Paris, France}

\author{Thomas Cope}
\affiliation{IQM Quantum Computers, Georg-Brauchle-Ring 23-25, 80992 Munich, Germany}

\author{Olivier Salomon}
\email{olivier.salomon@allianz.fr}
\affiliation{Allianz Quantum Hub, Paris, France}

\begin{abstract}

Quantum kernel methods are leading candidates for a practical quantum advantage in machine learning, but assessing that potential requires two quantities usually reported separately: how well a kernel performs on the task, and how far its geometry departs from the classical kernels available for the same problem. We introduce a fully unsupervised, multi-objective protocol that optimises simultaneously the normalised pseudo discrepancy (NPD), a label-free proxy for anomaly detection quality, and the geometric difference (GD) to a tuned classical reference kernel, selecting models from the resulting Pareto front. We apply it to malware beaconing detection in real network traffic, using a one-class support vector machine with fidelity and projected quantum kernels over four data encodings, on simulators and on IQM's 20-qubit Garnet processor. NPD-guided selection alone finds a fidelity kernel that beats the tuned classical baseline, but with a geometric difference too small to certify the gain as quantum. Projected kernels reach far larger geometric differences; the Pareto-selected one only marginally exceeds the baseline (AUC $0.782$ versus $0.765$, $g_{C\to Q}\approx 89>\sqrt{N}$ relative to that reference kernel), still below the NPD-selected fidelity kernel ($0.840$).

\end{abstract}

\maketitle

\section{Introduction}
\label{sec:intro}
Quantum machine learning (QML) is one of the most active research areas in quantum computing, and a key space for the search of quantum advantage in the noisy intermediate-scale quantum (NISQ) era \cite{biamonte2017quantum, cerezo2022challenges, wang2024comprehensive}. Models based on variational quantum circuits have been proposed, such as quantum neural networks, but their applicability remains limited due to the barren plateau problem, an issue suffered generally by variational quantum algorithms \cite{larocca2025barren}. In contrast, quantum kernel methods offer an alternative approach in which a quantum circuit estimates a kernel matrix and the resulting kernel is used by a classically trained support vector machine~\cite{havlivcek2019supervised}. Here, we use such a quantum kernel within a classical one-class SVM, rather than a fully quantum SVM algorithm~\cite{rebentrost2014quantum}. Quantum kernels use quantum circuits to define data-dependent similarity measures whose induced geometry may differ from that of the classical kernel families considered for a given task, potentially providing a useful task-dependent inductive bias.

Despite their promise, quantum kernels have yet to demonstrate a practical advantage. Furthermore, they present a fundamental trade-off in expressivity. A highly expressive kernel exploring a large Hilbert space can easily lead to an exponential concentration of fidelity kernel values, which vanish for most data points as the number of qubits increases \cite{thanasilp2024exponential, kairon2025equivalence}; the very feature hypothesized to provide advantage thus leads, when scaled up, to an essentially useless model.

This can be mitigated by optimizing the kernel bandwidth, as shown in \cite{shaydulin2022importance}, or by using projected quantum kernels (PQK), introduced in~\cite{huang2021power}, where the embedding quantum state is projected back to an approximate classical representation before computing the kernel. At the opposite extreme, however, a quantum kernel with low expressivity may not be distinguishable at all from a classical kernel. The data-dependent notion of geometric difference (GD) was introduced in \cite{huang2021power} to quantify the separation between classical and quantum kernels, and gives a recipe to construct artificial learning problems that best separate the corresponding models. We distinguish this structural separation from predictive quantum advantage: a large GD indicates that a quantum kernel is not well approximated, in the relevant geometric sense, by the chosen classical reference kernel, but it does not guarantee superior task performance. Prior works have tuned quantum kernels for predictive performance~\cite{hubregtsen2022training} or characterized their potential advantage over classical kernels through their spectral properties~\cite{kubler2021inductive}, while label-free hyperparameter selection, in which an internal quality proxy computed from the data replaces a labeled validation set, has been developed only for classical detectors: Dai and Fan introduced the normalized pseudo discrepancy (NPD) as such a proxy for unsupervised anomaly detection~\cite{dai2024autouad}. To our knowledge, no prior work has jointly optimized predictive performance and separation from classical kernels under a fully unsupervised protocol.

We address this gap with a multi-objective optimization framework that simultaneously maximizes the normalized pseudo discrepancy (NPD)~\cite{dai2024autouad}, a label-free anomaly detection quality proxy, and the GD to the reference classical kernel. This provides a systematic framework for balancing predictive performance and geometric separation. To showcase the application of the method in a realistic scenario, we test it on the unsupervised anomaly detection problem on real malware beaconing data -- the periodic communication between an infected host and a command-and-control server used to receive commands or exfiltrate data. We explore different fidelity kernels and PQKs, computed on noisy simulators and IQM's quantum hardware. Our results show that some projected quantum kernels achieve a large geometric difference (formalized in Section~\ref{sec:pareto}), indicating that a learning problem with potential quantum prediction advantage can be constructed from the data. However, this geometric separation brings at most a marginal performance gain on the malware detection task considered here.

The choice of malware beaconing as a proof of concept is deliberate, as the application of unsupervised quantum kernels to real cybersecurity data remains underexplored. The scarcity of malware data makes unsupervised models a requirement, while data fragmentation across the industry due to privacy and security concerns motivates the search for models that can identify complex patterns in limited data, a regime in which quantum kernels are more likely to generalize from few samples and separate from classical kernels~\cite{huang2021power,caro2022generalization}.

Our contributions can be summarized as follows:
\begin{enumerate}[label=(\roman*)]
\item A multi-objective (NPD + GD) Pareto framework for systematically searching for empirical, but not provable, quantum advantage in kernel methods, applicable beyond this use case.
\item NPD as a fully label-free criterion for quantum kernel hyperparameter tuning.
\item The first unsupervised quantum kernel benchmark on real malware beaconing data.
\item Empirical characterization of the performance--separability trade-off: fidelity kernels achieve the strongest detection performance while remaining below the geometric-difference scale beyond which classical kernels are no longer guaranteed to match quantum performance, whereas projected kernels exceed this scale without a corresponding gain in detection.
\item Hardware deployment on the 20-qubit IQM Garnet processor with simulator-vs-QPU comparison, showing that, at current noise levels, neither kernel family yet reaches the classical baseline.

\end{enumerate}

The rest of the paper is organized as follows. Section II discusses related work and background relevant to the results presented here. Section III describes all the methods, particularly the multi-objective kernel selection based on Pareto optimality and the feature engineering procedure. Section IV describes our results, including classical baselines, quantum kernels run on noisy simulators, and quantum kernels run on quantum hardware. We finish with a discussion of the results and our conclusions in Section V.


\section{Related Work}
\label{sec:related}

The question of whether quantum kernels offer a genuine advantage over classical ones predates our work and remains unsettled. On the positive side, rigorous constructions show that quantum kernels can yield a provable speedup for carefully designed learning problems~\cite{liu2021rigorous, muser2024provable}, and the geometric difference of Huang et al.~\cite{huang2021power} provides the theoretical lens through which such separations are usually argued. On the negative side, the picture on realistic data is far less encouraging. Fidelity kernels are prone to exponential concentration as expressivity grows~\cite{thanasilp2024exponential}, and systematic numerical comparisons repeatedly find that classical kernels match or exceed their quantum counterparts on standard benchmarks~\cite{slattery2023numerical}, while analyses of their generalization in the NISQ era further temper expectations~\cite{wang2021towards}. Even where a large geometric difference is present, it is necessary but not sufficient for advantage: the induced feature space must also carry an inductive bias aligned with the task at hand~\cite{kubler2021inductive}. This literature motivates our stance of treating geometric separation and predictive quality as two distinct criteria that must be assessed together, rather than assuming that one implies the other.

A second body of work concerns unsupervised anomaly detection and, in particular, the difficulty of model selection when labels are unavailable. One-class support vector machines~\cite{scholkopf2001support}, isolation forests~\cite{liu2008isolation}, and reconstruction-based autoencoders~\cite{hinton2006reducing} are long-standing baselines that are normally used as classical references. In all of them, performance is highly sensitive to hyperparameters, including the threshold used to flag anomalies, yet the usual practice of tuning against a labeled validation set is precisely what an unsupervised setting forbids. Label-free surrogates have therefore been proposed to guide model selection; among these, the normalized pseudo discrepancy (NPD) of Dai and Fan~\cite{dai2024autouad} measures the separation between anomaly scores on held-out normal data and on synthetic out-of-distribution samples, and correlates with detection quality without any access to labels. Such surrogates, however, have so far guided only classical model selection.

Applications of quantum machine learning to cybersecurity, finally, remain comparatively sparse, and the existing efforts fall into categories that do not overlap with the regime we study. A first line addresses supervised classification of malware and network attacks, comparing quantum and classical models where labels are abundant~\cite{barrue2023quantum, payares2021quantum, kumar2025quids, cultice2024anomaly}, or relaxes the requirement only to the semi-supervised case~\cite{tscharke2023semisupervised}. A second line does operate without labels, building on quantum autoencoders for network-traffic anomaly detection~\cite{hdaib2024quantum}; such reconstruction-based pipelines are not organized around a quantum kernel, and so do not expose the kernel matrix that the geometric difference requires. A third line applies the combination of quantum kernels and unsupervised anomaly detection to non-cybersecurity (financial and physics) data~\cite{pranjic2024unsupervised}, and evaluates detection across a controlled range of anomaly ratios, treating the contamination level as a known quantity rather than unknown, as required in an operational deployment. None of the above works combines all three ingredients at once: an unsupervised protocol, a quantum kernel amenable to geometric analysis, and real cybersecurity data. This intersection is where malware beaconing detection sits, and where labeled data is genuinely scarce and quickly rendered obsolete by concept drift~\cite{gama2014conceptdrift, shalaginov2016malware}, so that a fully unsupervised treatment is not a convenience but a requirement. Our work is, to our knowledge, the first to occupy it.


\section{Methods}
\label{sec:methods}

\subsection{Kernel methods}
\label{sec:quantum-kernels}
Kernel methods are widely used in machine learning to solve, e.g., classification and regression problems. The key idea of these techniques is to map data, by means of a non-linear feature map, from an input space to a (usually higher dimensional) feature space, where the learning task becomes easier to analyze, i.e.\ linear. The data can then be processed in the feature space by accessing solely the kernel values, which represent a similarity measure between data, without explicitly computing the feature map. This is usually referred to as the \textit{kernel trick}.

We now define the notation for the quantum kernels used in the following sections~\cite{schuld2021supervised}. A kernel $k: \mathcal{X}\times\mathcal{X}\rightarrow\mathbb{R}$ is a positive definite function in two data points $x,x'$. Let $\phi:x \rightarrow \rho(x)$ be a data-encoding feature map over the input space $\mathcal{X}$, where $\rho$ is a density matrix. A quantum kernel can then be defined as the inner product of the feature vectors:
\begin{equation}
  k(x, x') = \mathrm{Tr}\left(\rho(x')\rho(x)\right) = \left| \langle \phi(x') \mid \phi(x) \rangle \right|^2. \label{eq:k}
\end{equation}
where the second equality holds if the data-encoding quantum state is pure. We will refer to the kernel in Eq.~\eqref{eq:k} as the \textit{fidelity kernel}. The second kernel considered in this work is the projected quantum kernel introduced in~\cite{huang2021power}: 
\begin{equation}
  k(x,x') = \exp\left(-\gamma \sum_{k} \lVert\rho_k(x)-\rho_k(x')\rVert_F^2\right)
  \label{eq:pqkdef}
\end{equation}
where $\gamma>0$ is a hyperparameter, $\lVert\cdot\rVert_F$ is the Frobenius norm and $\rho_k(x)=\mathrm{Tr}_{j\neq k}(\rho(x))$ is the single-qubit reduced density matrix. The hyperparameter $\gamma$ acts as an inverse lengthscale in the space of reduced density matrices, setting the distance over which two encoded states count as similar; it plays the role of the bandwidth of a classical RBF kernel, whose tuning is known to mitigate exponential kernel concentration~\cite{shaydulin2022importance}.\\
Given the kernel function, for a finite set of data points we can define the kernel matrix as the square matrix whose entries are given by the inner product between all possible pairs. This matrix is then fed to a classical anomaly detection algorithm. In our case, we use a One-Class Support Vector Machine (OCSVM) which learns a decision boundary that encloses normal data given a kernel matrix~\cite{bounsiar2014one}.
In the quantum case, the data-encoding feature map is a unitary $U(x)$ implemented by a quantum circuit. Previous works have studied how to design classically difficult feature maps to get a provable advantage over classical kernels~\cite{liu2021rigorous} or have identified a promising class of learning problems for quantum kernels~\cite{glick2021covariant}. As the focus of this work is to present a fully unsupervised pipeline for anomaly detection with quantum kernels, we do not engineer a new feature map, but use encodings already present in the literature. The choice of the optimal encoding becomes part of the multi-objective optimisation presented in Sec.~\ref{sec:pareto} below. We consider four different data-encoding circuits, whose definition and details can be found in Appendix \ref{sec:supplementary}.

\subsection{Multi-objective quantum kernel selection}
\label{sec:pareto}

The main methodological contribution of this work is the development of a fully unsupervised quantum kernel selection that simultaneously optimizes quantum kernel performance via the NPD score and separation with respect to classical kernels via the GD.

Starting with the quantum kernel performance, the absence of labels in unsupervised settings poses a challenge for hyperparameter tuning of ML models. To address this issue, we use the normalized pseudo-discrepancy (NPD) score, first introduced in Ref.~\cite{dai2024autouad}, which measures the discrepancy between the anomaly scores of a validation set drawn from the training data and a validation set drawn from an axis-aligned Gaussian fitted to the training data.

More precisely, given an unsupervised anomaly detection model $M$ with hyperparameters $\Theta$, the training dataset $X$ is randomly split into a training subset $X_{\mathrm{trn}}$ and a validation subset $X_{\mathrm{val}}$, where $|X_{\mathrm{val}}| = m$. The model is trained only on $X_{\mathrm{trn}}$. Let $\boldsymbol{\mu}_{\mathrm{trn}} \in \mathbb{R}^d$ and $\boldsymbol{\sigma}^2_{\mathrm{trn}} \in \mathbb{R}^d$ denote the empirical mean and variance vectors of $X_{\mathrm{trn}}$, respectively. A generated dataset $X_{\mathrm{gen}}$ of size $m$ is then sampled from an
axis-aligned Gaussian distribution with diagonal covariance
\[
X_{\mathrm{gen}} \sim \mathcal{N}\!\left(\boldsymbol{\mu}_{\mathrm{trn}},
\operatorname{diag}(\boldsymbol{\sigma}^2_{\mathrm{trn}})\right),
\]
whose per-dimension variances match those of $X_{\mathrm{trn}}$.

Let $s_{\mathrm{val}} = f_M(X_{\mathrm{val}} \mid \Theta)$ and $s_{\mathrm{gen}} = f_M(X_{\mathrm{gen}} \mid \Theta)$ denote the anomaly score vectors produced by the model for the validation and generated samples, respectively. The NPD metric is then defined as
\[
V_{\mathrm{NPD}}(M, X)
=
\frac{\left(\operatorname{Mean}(s_{\mathrm{gen}}) - \operatorname{Mean}(s_{\mathrm{val}})\right)^2}
{2\left(\operatorname{Var}(s_{\mathrm{gen}}) + \operatorname{Var}(s_{\mathrm{val}})\right) + \varepsilon},
\]
where $\varepsilon > 0$ is a small constant to ensure numerical stability.

Intuitively, NPD measures how well the model separates the validation samples, assumed to be predominantly normal, from synthetically generated samples that are more diverse and likely to lie outside the normal data manifold. A larger NPD indicates a clearer separation in anomaly scores, which empirically correlates with better detection performance on unseen test data. Importantly, NPD is hyperparameter-free, translation- and scale-invariant with respect to the scoring function, and mitigates overfitting by relying on data ($X_{\mathrm{val}}$ and $X_{\mathrm{gen}}$) that are independent of model training.

On the other hand, the potential advantage offered by a quantum model over a classical model can be evaluated by means of the geometric difference, as introduced in Ref.~\cite{huang2021power} and defined as

\begin{equation}
g_{C \rightarrow Q}
= \sqrt{\left\lVert \sqrt{K_Q}\, K_C^{-1}\, \sqrt{K_Q} \right\rVert_{\infty} },
\label{eq:geometric-difference}
\end{equation}

where $K_C$ and $K_Q$ denote the classical and quantum kernel matrices, respectively, with $K_C$ taken as the kernel of the tuned classical OCSVM baseline, both normalised such that $\mathrm{Tr}(K_Q) = \mathrm{Tr}(K_C) = N$. The notation $\lVert \cdot \rVert_{\infty}$ represents the matrix spectral norm. Intuitively, if $K_C(x_i, x_j)$ is small (large) whenever $K_Q(x_i, x_j)$ is small (large), then the geometric difference takes a small value. A small geometric difference indicates that the classical model is guaranteed to achieve performance similar to, or better than, the quantum model. Conversely, the geometric difference increases as the two kernels deviate from one another.

Furthermore, it is shown in~\cite{huang2021power} that the classical model is expected to provide similar or better performance when
\begin{equation}
g_{C \rightarrow Q} \ll \sqrt{N},
\label{eq:GD}
\end{equation}
where $N$ is the number of samples in the training dataset.

With both objectives defined, we search over feature maps and OCSVM hyperparameters for the Pareto front, i.e.\ the set of non-dominated solutions: a solution is non-dominated when no other candidate is at least as good in both objectives and strictly better in one, so neither NPD nor GD can be improved without degrading the other. We apply this approach to both the fidelity and the projected quantum kernels. Table \ref{tab:quantum-kernel-hyperparameters} gives the list of hyperparameters that were optimised. To obtain the Pareto front, we used the open-source package Optuna~\cite{akiba2019optuna} in addition to the NSGA-II sampler~\cite{deb2002fast}. NSGA-II is a genetic multi-objective optimisation algorithm. It begins from an initial population of candidate solutions sampled from the search space and evolves it over successive generations, evaluating each candidate on both objectives and carrying the best ones forward. Here, a candidate solution is a choice of hyperparameters from the search space in Table~\ref{tab:quantum-kernel-hyperparameters}. From the Pareto frontier of the final population, we select as the best solution the knee point, i.e.\ the best-compromise configuration: both objectives are rescaled to $[0,1]$ across the front, and we retain the solution with the smallest Euclidean distance to the utopia point $(1,1)$ at which both NPD and GD would be simultaneously maximal. This avoids the extremes of the front, where one objective is optimised at the expense of the other. 

\begin{figure}
    \centering
    \includegraphics[width=\columnwidth]{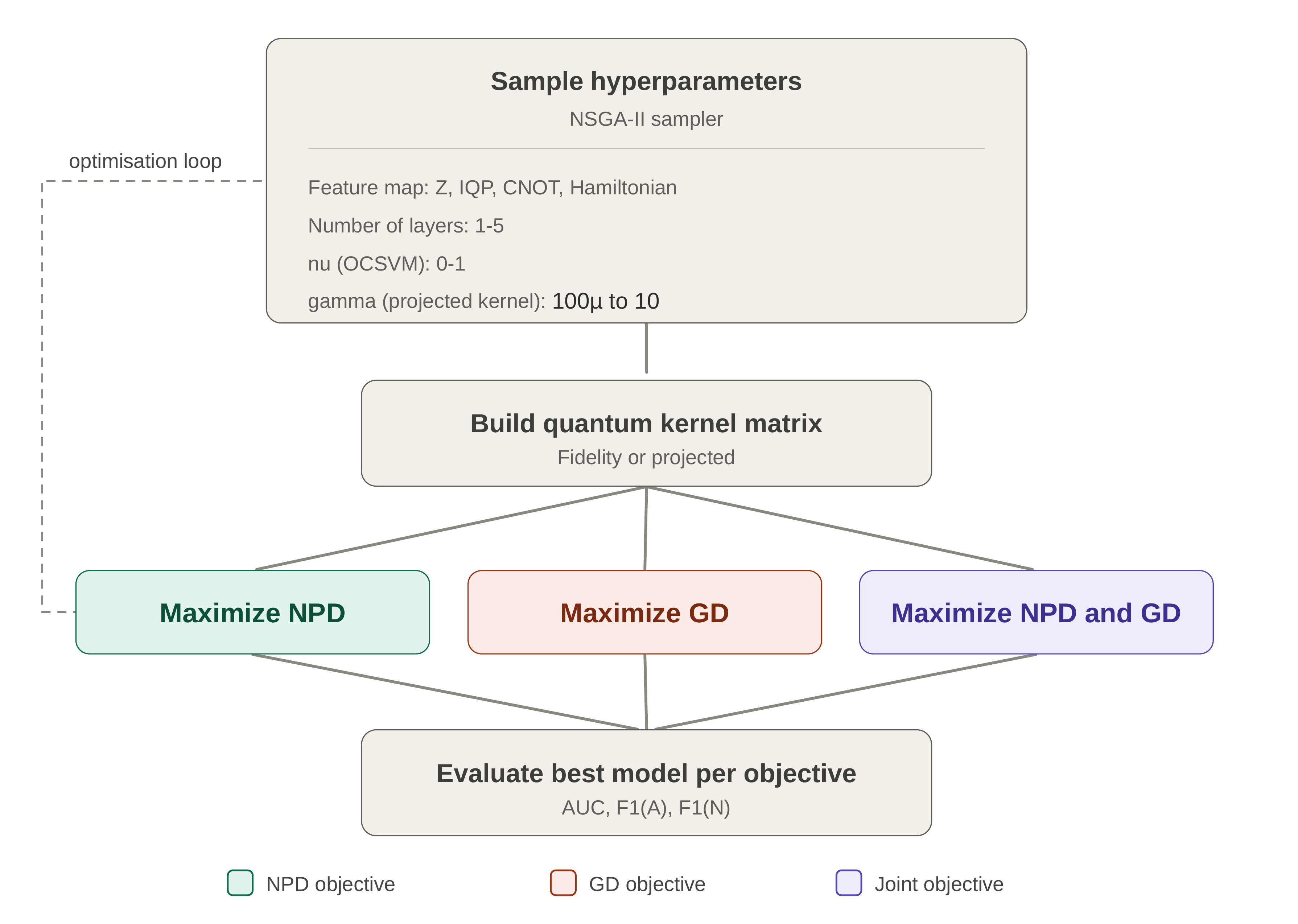}
    \caption{The unsupervised optimisation pipeline: a hyperparameter search over quantum kernels selected under three objectives (NPD, GD, and their joint Pareto knee) and evaluated on the test set.}
    \label{fig:optimisation}
\end{figure}

In addition to the joint multi-objective optimisation of the NPD and GD scores, we also optimised each objective independently. For these single-objective runs, Optuna defaults to the Tree-structured Parzen Estimator (TPE)~\cite{watanabe2023tree}, a Bayesian sampler that models the promising and unpromising regions of the search space separately and proposes new configurations by maximising their density ratio. This yields three sets of optimised hyperparameters per kernel—one maximising NPD alone, one maximising GD alone, and one balancing both. Figure \ref{fig:optimisation} shows a summary of the three optimisations. 

\begin{table*}[t]
\centering
\small
\begin{ruledtabular}
\begin{tabular}{lll}
\textbf{Model} & \textbf{Hyperparameter} & \textbf{Search Space} \\
\colrule

Projected Quantum Kernel (RBF)
& Feature map
& Z, IQP, CNOT, Hamiltonian \\

& Number of layers
& $\{1,2,3,4,5\}$ \\

& K-RDM
& $\{1\}$ \\

& Gamma ($\gamma$)
& $[100\mu,10]$ \\

& $\nu$
& $[0,1]$ \\

\colrule

Fidelity Kernel
& Feature map
& Z, IQP, CNOT, Hamiltonian \\

& Number of layers
& $\{1,2,3,4,5\}$ \\

& $\nu$
& $(0,1]$ \\

\end{tabular}
\end{ruledtabular}

\caption{Hyperparameter search space for quantum kernel methods. The description of the considered feature maps is given in Appendix \ref{sec:supplementary}. The number of layers ranged from 1 to 5. We only considered the single-qubit reduced density matrix for the projected quantum kernels, and $\gamma$ is the hyperparameter of the projected quantum kernel. $\nu$ is a hyperparameter of the OCSVM model, which represents an upper bound on the fraction of training errors and a lower bound of the fraction of support vectors. 
Curly brackets $\{\}$ denote discrete values, while square brackets
$[]$ denote continuous ranges.}
\label{tab:quantum-kernel-hyperparameters}
\end{table*}

\subsection{Dataset}
\label{sec:data}

To test our systematic approach of finding the optimal quantum kernel model, we decided to test it on a real-world malware beaconing dataset. Malware beaconing represents periodic communication between an infected host and an external C\&C (command-and-control) server. These signals are difficult to detect as they can easily blend in with benign network traffic. As such, a high-quality dataset is essential for network traffic analysis. The Stratosphere Research Laboratory is dedicated to capturing real malware traffic and providing open-access network traffic datasets to support the development of robust machine learning models for malware detection~\cite{Stratosphere_normal}. We first obtain normal network traffic datasets from their platform to define what it is normal for a typical user. 

The laboratory also provides the CTU-13 dataset, which is a beacon network traffic dataset consisting of large-scale captures of real botnet activity mixed with normal and background traffic. The dataset comprises thirteen scenarios, each corresponding to a different botnet sample. In each scenario, a specific malware instance was executed, using multiple protocols and performing diverse actions. For each capture, the laboratory provides a list of external control IP addresses, which enabled us to filter out normal and background traffic from the botnet-related traffic.

Since we are focused on unsupervised learning, the training dataset only contains normal network traffic. The beacon traffic is all kept in the test dataset.

Network traffic was aggregated into unique flows (source IP address, destination IP address, protocol and external port number), each described by a set of statistical features. From this initial set, we retained a non-redundant subset of $6$ features, given in Table~\ref{tab:features}; the full feature list and the selection procedure, which is unsupervised and uses no label information, are detailed in Appendix~\ref{sec:features}. The training set contains $200$ randomly sampled normal flows, while the testing set contains $1{,}000$ flows, of which 457 are anomalous flows and 543 are normal flows. All malicious flows are included exclusively in the testing set. For the simulator results, detection scores are aggregated per unique source--destination pair, keeping the lowest score of each pair, which gives $777$ evaluation samples ($255$ anomalous and $522$ normal); the hardware results are evaluated on the $1{,}000$ flows.

\begin{table}[h!]
\centering
\caption{Feature set obtained from the feature engineering procedure}
\label{tab:features}
\begin{tabular}{c}
\toprule
\textbf{Feature Name} \\
\midrule
Forward packets per second \\
Backward packets per second \\
Standard deviation of the FFT of forward traffic \\
Standard deviation of forward inter-arrival time \\
Standard deviation of packet lengths \\
Mean forward packet length \\

\bottomrule
\end{tabular}
\end{table}


\section{Results}
\label{sec:results}
To evaluate the performance of the models, we use two metrics: the area
under the ROC curve (AUC) and the F1 score. The AUC measures the model's
ability to rank anomalies above normal instances across all decision
thresholds, while the F1 score is the harmonic mean of precision and
recall at a fixed threshold. Because the F1 score depends on which class
is designated positive, we report it twice, treating anomalies (A) and normal
instances (N) as the positive class in turn. In our view, reporting both
provides a balanced and transparent view of model performance.

\subsection{Classical baselines}
\label{sec:res-classical}
Here, we present the performance of the classical OCSVM on the malware dataset.  Tuning the model to maximise the NPD score selects an RBF kernel with $\gamma = 8.99$ and $\nu = 0.472$.

\begin{table}[ht]
\centering
\caption{Performance of the classical OCSVM on the malware dataset.}
\label{tab:classical_baseline}
\begin{tabular}{ccc}
\toprule
AUC & F1 (A) & F1 (N) \\
\midrule
0.765 & 0.599 & 0.561 \\
\bottomrule
\end{tabular}
\end{table}

The performance of this configuration on the test set is summarised in Table~\ref{tab:classical_baseline}. The classical OCSVM attains an AUC of $0.765$, indicating a moderate ability to rank anomalous flows above normal ones. The F1 scores are more modest: $0.599$ with anomalies as the positive class and $0.561$ with normal flows as the positive class.  Taken together, these results establish a non-trivial baseline against which we assess the quantum kernels.

\subsection{Fidelity and projected quantum kernels: NPD, GD, and Pareto optimization}
\label{sec:res-fidelity}
For the quantum kernel models, we performed a hyperparameter search to identify the optimal quantum kernel. We are interested in three objectives: (1) the NPD score, (2) GD between the quantum kernel and the reference classical kernel, and (3) a Pareto frontier that jointly maximises both the NPD and GD.
 
Table \ref{tab:quantum-kernel-hyperparameters} gives the hyperparameter search space for both the fidelity and projected quantum kernels. For each optimisation objective, we conducted 100 trials of hyperparameter search (150 for the fidelity kernel). The best quantum model for each objective was then evaluated on the test dataset. Using the statevector simulator, Tables \ref{tab:best_config_fidelity} and \ref{tab:best_config_pqk} present the best configurations chosen for the three objectives for the fidelity and projected quantum kernels, respectively.

\begin{table}[t]
    \centering
    \caption{Best fidelity quantum kernel configuration found for the three objectives on the dataset using the statevector simulator. The GD objective depends only on the kernel matrix, so $\nu$ is not defined for that column (--).}
    \label{tab:best_config_fidelity}
    \begin{tabular}{lccc}
        \toprule
        & NPD & GD & Pareto \\
        \midrule
        Feature map & Hamiltonian & Hamiltonian & Hamiltonian \\
        Layers & 2 & 1 & 5 \\
        $\nu$ & 0.322 & -- & 0.266 \\
        \midrule
        $g_{C \rightarrow Q}$ & 4.297 & 7.14   & 5.94  \\
        NPD & 2.34 & 0.592 & 0.72 \\
        \bottomrule
    \end{tabular}
\end{table}

\begin{table}[t]
    \centering
    \caption{Best projected quantum kernel configurations found for the three objectives on the six features dataset using the statevector simulator. The GD objective depends only on the kernel matrix, so $\nu$ is not defined for that column (--).}
    \label{tab:best_config_pqk}
    \begin{tabular}{lccc}
        \toprule
        & NPD & GD & Pareto \\
        \midrule
        Feature map & Z & Hamiltonian & IQP-style \\
        Layers & 5 & 1 & 5 \\
        $\nu$ & 0.385 & -- & 0.104 \\
        $\gamma$ & 2.37 & 9.99 & 0.000395 \\
        \midrule
        $g_{C \rightarrow Q}$ & 32.0 & 272  & 88.6 \\
        NPD & 2.94 & 0.603 & 1.42 \\
        \bottomrule
    \end{tabular}
\end{table}

 From Table \ref{tab:best_config_fidelity}, the Hamiltonian feature map is selected across all three objectives for the fidelity quantum kernel. Maximising the geometric difference (GD) yields a maximum GD of $7.14$. This demonstrates that the fidelity quantum kernel achieves only small geometric differences (GD) on this dataset. Hence, even when the fidelity quantum kernel models perform well, a classical model can match its performance due to these consistently small GD values across all three objectives.

We observe a different behaviour, however, for the projected quantum kernel, as shown in Table \ref{tab:best_config_pqk}. Maximising the GD yields a value of $272$, which is greater than $\sqrt{N} = 14.14$, demonstrating that, at least for this dataset, the projected quantum kernel can lead to new kernels that are distinct from the reference classical kernel (the NPD-tuned RBF).

At first sight it may seem counter-intuitive that the projected kernels, which discard all but the single-qubit marginals of the encoded state, are the ones reaching large GD values; the same effect was reported in Ref.~\cite{huang2021power}, where the GD was introduced. The reason is that GD measures how well the classical kernel matrix can reproduce the \emph{geometry} of the quantum one, not how much quantum information the kernel retains. The fidelity kernel has no hyperparameter beyond the encoding circuit itself: its entries are state overlaps $\mathrm{Tr}(\rho(x')\rho(x))$, and for the shallow circuits considered here the resulting similarity structure remains close to that of a smooth classical kernel on the same six features, which bounds the attainable GD. The projected kernel instead composes the encoding with an RBF kernel of tunable bandwidth $\gamma$ acting on the reduced density matrices, so the optimiser can reshape the spectrum of $K_Q$ directly. Large $\gamma$ drives $K_Q$ towards the identity: the diagonal is always one, while the off-diagonal entries vanish, so no pair of distinct points is regarded as similar. This is the regime in which the reference kernel $K_C$, whose bandwidth was tuned for NPD and which retains non-trivial off-diagonal structure, lies furthest from $K_Q$: for $K_Q = I_N$ one has $g_{C\to Q}^2=\lVert K_C^{-1}\rVert_\infty$, which is set by the smallest eigenvalue of $K_C$. The GD-maximising configuration indeed sits at the top of the $\gamma$ range ($\gamma = 9.99$, $g_{C\rightarrow Q} = 272$). A large GD is therefore a property of the pair $(K_Q,K_C)$ and not evidence of quantum structure: a classical RBF kernel with a larger bandwidth parameter would reproduce the same identity-like geometry. The same narrow-bandwidth limit is the memorising regime in which generalisation deteriorates~\cite{shaydulin2022importance}.

\begin{figure*}
    \centering
    \includegraphics[width=\textwidth]{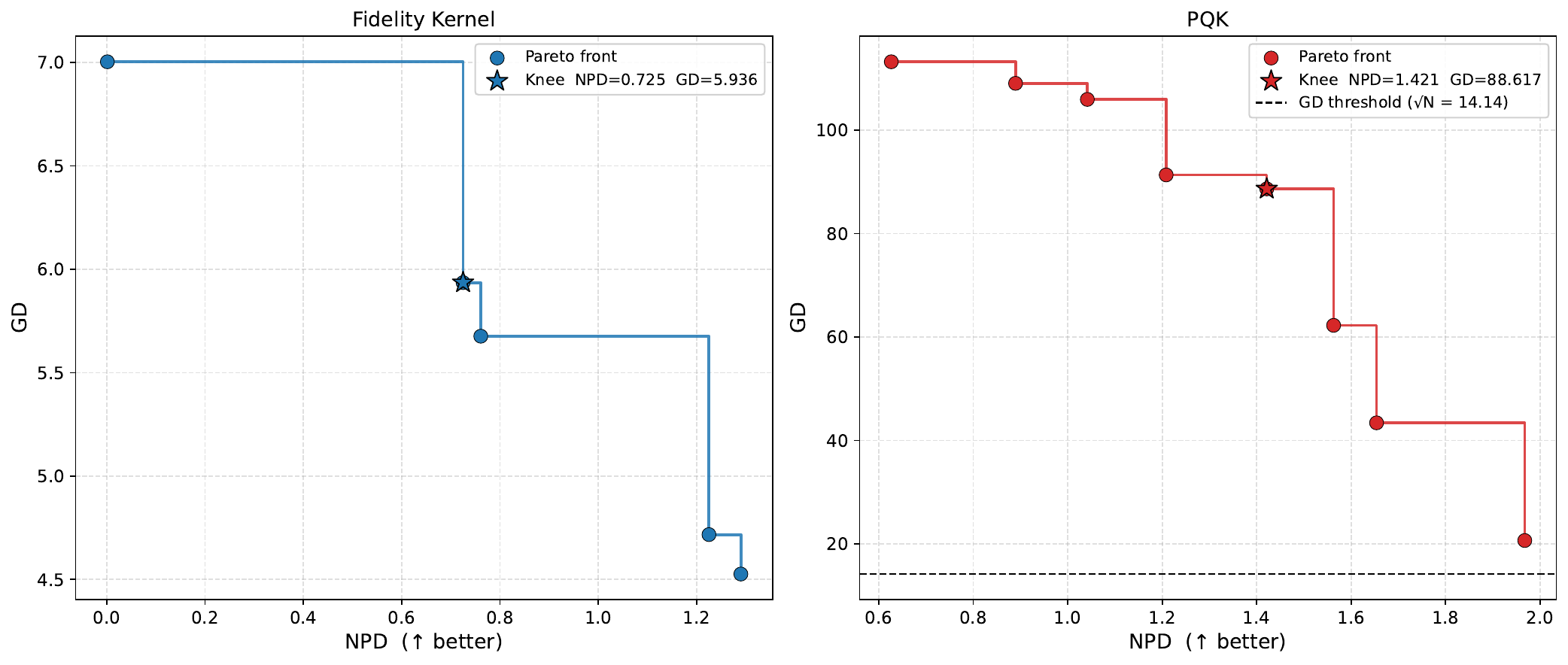}
    \caption{Pareto fronts of the fidelity (blue) and projected (red)
    quantum kernels under joint maximisation of the NPD and GD, evaluated on
    the statevector simulator. Circles mark the non-dominated configurations
    of each front; stars mark the selected knee points. The fidelity kernel maintains a low GD across the front, whereas the
    projected kernel trades a high GD at low NPD for improved NPD, exposing
    the tension between the two objectives. Each front contains only the configurations that remained non-dominated during the optimisation, which is why the two kernels contribute different numbers of points.}
    \label{fig:pareto_front}
\end{figure*}

Figure~\ref{fig:pareto_front} shows the Pareto fronts obtained when the
NPD and GD are maximised jointly, for both the fidelity and projected
quantum kernels on the statevector simulator. For the fidelity kernel,
the GD stays low throughout
($\mathrm{GD}\lesssim 7$), decreasing as the NPD
increases, with the knee at $\mathrm{NPD}=0.725$,
$\mathrm{GD}=5.936$. This indicates that, at least for this dataset,
the fidelity kernel never leaves the regime $\mathrm{GD}\ll\sqrt{N}$ in
which a classical kernel is guaranteed to match its performance. The
projected quantum kernel behaves very differently. Its front is a
broad, monotonic trade-off: the GD reaches $\approx 110$ at low NPD and
falls as the NPD improves, passing through the knee at
$\mathrm{NPD}=1.421$, $\mathrm{GD}=88.6$, down to
$\mathrm{GD}\approx 20$ at $\mathrm{NPD}\approx 2$. 
Even the front with the lowest GD  remains above the threshold, so every configuration along the projected
front lies in the regime where the reference classical kernel is not guaranteed to reproduce its geometry.

\begin{figure}
    \centering
    \includegraphics[width=\columnwidth]{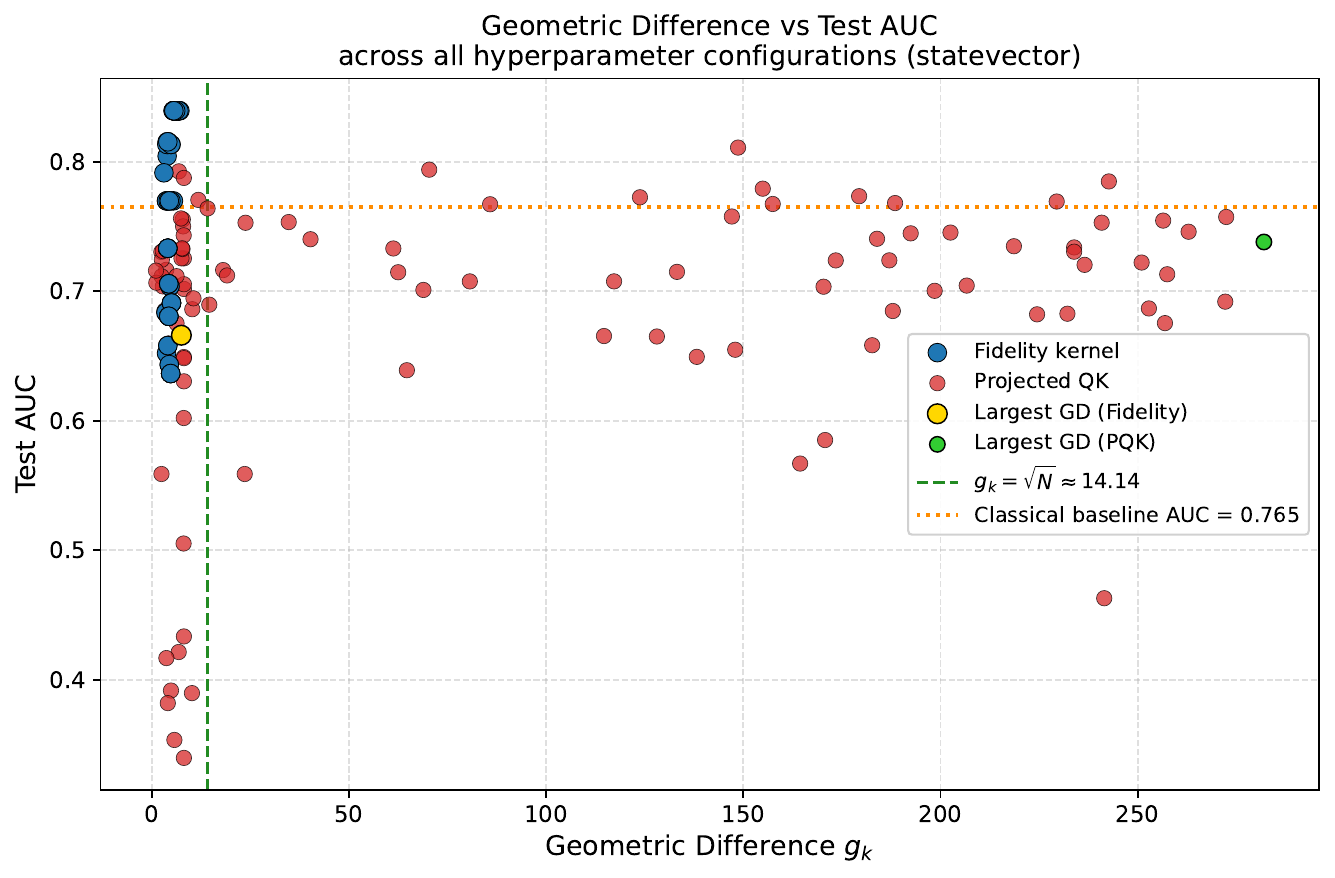}
\caption{Test AUC versus geometric difference $g_k$ for all
hyperparameter configurations evaluated during the GD-maximisation run
(statevector simulation). Blue: fidelity kernels; red: projected
quantum kernels. The green dashed line marks $g_k=\sqrt{N}\approx
14.14$, below which a classical kernel is guaranteed to match the
quantum model; the orange dotted line is the tuned classical OCSVM
baseline (AUC $=0.765$). Fidelity kernels stay below the threshold yet
reach the highest AUC in the run ($\approx 0.84$), whereas projected
kernels extend to $g_k\sim\!270$ with an AUC that shows no systematic
trend with $g_k$. A few projected configurations combine
$g_k\gg\sqrt{N}$ with an AUC above the baseline, but none reaches the
best fidelity kernels, and the two largest-GD models (yellow, green)
perform at or below the baseline.}
    \label{fig:gd_auc}
\end{figure}

\begin{figure}
    \centering
    \includegraphics[width=\columnwidth]{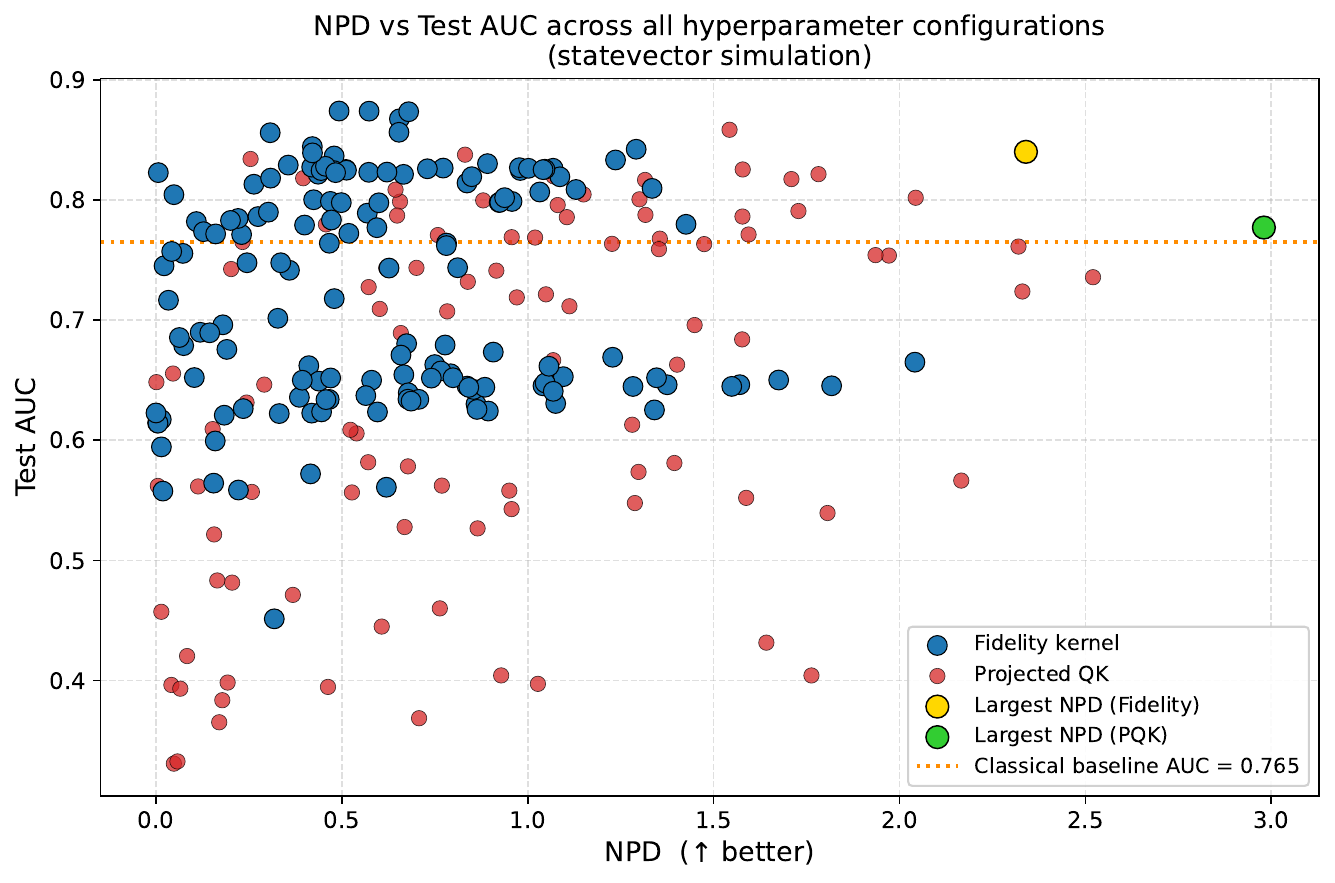}
\caption{Test AUC versus NPD for all hyperparameter configurations
evaluated during the NPD-maximisation run (statevector simulation).
Blue: fidelity kernels; red: projected quantum kernels. The orange
dotted line marks the tuned classical OCSVM baseline (AUC $=0.765$).
Fidelity kernels concentrate at higher AUC, with a substantial fraction
above the baseline, while projected kernels spread widely between AUC
$\approx 0.33$ and $0.86$ and mostly lie below it. Within each family
the NPD--AUC relation is weak. The largest-NPD configuration of each
family (yellow: fidelity; green: projected) lies above the baseline, at
AUC $=0.840$ and $0.777$ respectively.}
    \label{fig:npd_auc}
\end{figure}

Table~\ref{tab:benchmark} reports the test performance of the configurations
selected under each objective. Selecting for geometric difference neither
helps nor substantially hurts detection: the GD-maximising projected kernel,
which attains the largest GD in our study ($g_{C\rightarrow Q}=272$), reaches
an AUC of $0.738$, close to the classical baseline ($0.765$) and to the
NPD-selected projected kernel ($0.777$, $g_{C\rightarrow Q}=32$), despite an
almost tenfold larger GD. The best model overall is instead the NPD-selected
fidelity kernel, which outperforms the classical baseline with an AUC of
$0.840$ at a GD of just $4.297$. This shows that, at least for this dataset, a large GD does not translate into better performance. Figures~\ref{fig:gd_auc} and~\ref{fig:npd_auc} examine this relation across all explored configurations.

Figure~\ref{fig:gd_auc} plots the test AUC of every configuration
explored during the GD-maximisation run against its geometric difference
$g_k$. The fidelity kernels (blue) stay confined to small $g_k$, below the $g_k=\sqrt{N}\approx 14.14$ threshold (green dashed), and
reach the highest AUC in the run ($\approx 0.84$), with multiple configurations also above the classical baseline (orange dotted). The projected
kernels (red) extend to $g_k\sim\!270$, far beyond the threshold, but
their AUC shows no systematic trend with $g_k$: beyond the threshold
they scatter between $\approx 0.46$ and $0.81$, and the poorest models
in the run (AUC $<0.5$) mostly sit at small $g_k$. A handful of projected
configurations do occupy the upper-right region, combining
$g_k\gg\sqrt{N}$ with an AUC above the baseline, yet none of them
reaches the best fidelity kernels. The extreme configurations make the
point sharply: the GD-maximising projected kernel (green) lands close to
the baseline, no better than configurations with $g_k$ ten times
smaller, while the GD-maximising fidelity kernel (yellow) is among the
weakest fidelity models at AUC $\approx 0.67$. Within the projected family, configurations with small GD generally
perform poorly, but increasing the GD does not reliably improve
detection. The best model overall is a fidelity kernel,
with a GD below $7$. This confirms that, on this dataset, a large GD is not required for good detection performance.

Figure~\ref{fig:npd_auc} shows the test AUC against the NPD for all
configurations of the NPD-maximisation run. The two families are
stratified: the
fidelity kernels (blue) concentrate between AUC $\approx 0.6$ and
$0.88$, with many configurations above the classical baseline
(orange dotted), while the projected kernels (red) are spread more
widely, from AUC $\approx 0.33$ to $0.86$, and mostly lie below it, so
the attainable performance here depends strongly on the kernel type.
Within each family the relation is weak: neither family shows a
systematic trend in AUC over $0 \lesssim \mathrm{NPD} \lesssim 2$.
The extremes nonetheless behave broadly as the criterion intends. The
largest-NPD fidelity configuration (yellow, AUC $0.840$) and the
largest-NPD projected configuration (green, AUC $0.777$) are both above the classical baseline,
although neither is the single best configuration of its family, with
a few models reaching AUC $\approx 0.87$ and $\approx 0.86$,
respectively. The NPD is therefore best understood as a criterion that
identifies a good region of the hyperparameter space rather than as a fine-grained
ranking of configurations. This behaviour is consistent with the
original study of NPD~\cite{dai2024autouad}: across 38 benchmark
datasets, NPD-guided Bayesian optimisation of a classical OCSVM came
close to the best attainable AUC ($84.0$ versus $85.8$, against $78.7$
with default hyperparameters) and outperformed the other label-free
criteria tested, while the authors note that the highest NPD did not
always correspond to the best model. Our results extend this picture to
quantum kernels. The distinction between identifying a good region and
ranking configurations within it is worth stating explicitly, since in
a fully unsupervised setting the NPD is the only selection signal
available; establishing how far label-free proxies track detection
performance is an open question that our results suggest deserves
dedicated study.

\begin{table*}[t]
\centering
\caption{Performance of the optimised quantum kernels obtained from the optimisation of the three objectives: NPD, GD and Pareto frontier using the statevector simulator.
We re-run the optimal hyperparameters selected by each optimisation using the IQM noisy simulator; the OCSVM parameter $\nu$ was additionally re-tuned on the noisy simulator by maximising the NPD. The classical OCSVM baseline achieves AUC $=0.765$.  The best value of each metric is shown in bold.}
\label{tab:benchmark}

\small
\setlength{\tabcolsep}{5pt}

\begin{tabular}{lccc|ccc|ccc}
\toprule
\textbf{Method}
& \multicolumn{3}{c|}{\textbf{NPD}}
& \multicolumn{3}{c|}{\textbf{GD}}
& \multicolumn{3}{c}{\textbf{Pareto (NPD \& GD)}} \\
\cmidrule(lr){2-4}
\cmidrule(lr){5-7}
\cmidrule(lr){8-10}

& AUC & F1(A) & F1(N)
& AUC & F1(A) & F1(N)
& AUC & F1(A) & F1(N) \\
\midrule

\multicolumn{10}{c}{\textbf{Statevector Simulator}} \\[2pt]

Fidelity
& \textbf{0.840} & 0.657 & 0.715
& 0.666 & 0.494 & \textbf{0.800}
& 0.812 & \textbf{0.671} &  0.773\\

Projected
& 0.777 & 0.599 & 0.563
& 0.738 & 0.542 & 0.340
& 0.782  & 0.638 & 0.736 \\

\midrule

\multicolumn{10}{c}{\textbf{IQM Noisy Simulator}} \\[2pt]

Fidelity
& 0.790 & 0.600 & 0.582
& 0.752 & 0.582 & \textbf{0.630}
& \textbf{0.798} & 0.600 & 0.576 \\

Projected
& 0.777 & \textbf{0.609} & 0.554
& 0.738 & 0.541 & 0.334
& 0.647 & 0.525 & 0.620 \\

\bottomrule
\end{tabular}
\end{table*}

Returning to the joint optimisation, the fronts in Fig.~\ref{fig:pareto_front}
should be read as the non-dominated set reached after a finite budget of
100 NSGA-II trials, not as converged approximations of the true Pareto
frontier. More tellingly, the single-objective TPE runs find
configurations outside the joint fronts. The NPD-maximising projected kernel
(NPD $= 2.94$, $g_{C\to Q} = 32.0$, Table~\ref{tab:best_config_pqk}) dominates the
high-NPD end of the projected front, where NPD $\approx 2$ and
$g_{C\to Q} \approx 20$, and the GD-maximising run reaches $g_{C\to Q} = 272$, above the
front's maximum of about 120. Likewise, the GD-maximising fidelity
configuration ($\mathrm{NPD} = 0.592$, $g_{C\to Q} = 7.14$,
Table~\ref{tab:best_config_fidelity}) dominates the low-NPD end of the fidelity
front. The joint search has therefore not fully explored either
extreme. However, the separation between the two kernel families
relative to $\sqrt{N}$ holds across all three optimisation runs and
does not depend on the front being converged.

We also train the selected models with the IQM noisy simulator, which
contains information about the IQM systems, like the number of qubits
and their connectivity, and several noise parameters (relaxation and
dephasing times, gate infidelities and readout errors). Comparing the
IQM noisy simulator results with the noiseless statevector ones reveals
that the impact of noise depends less on the kernel family than on the
circuit behind each configuration. The fidelity kernel degrades
moderately under NPD selection, with its AUC falling from $0.840$ to
$0.790$, and only slightly for the Pareto configuration, from $0.812$
to $0.798$, while the shallow GD configuration in fact improves, from
$0.666$ to $0.752$. The projected kernel is entirely unaffected under
NPD and GD selection, with AUCs of $0.777$ and $0.738$ in both
settings, but drops sharply for the Pareto configuration, from $0.782$
to $0.647$. As a result, the clear advantage the fidelity kernel holds
on the statevector simulator under NPD selection (AUC $0.840$ versus
$0.777$) narrows to a small margin under noise ($0.790$ versus
$0.777$). Under noise, the best model is the Pareto-selected fidelity
kernel ($0.798$), closely followed by the NPD-selected one, and both
remain above the classical baseline of $0.765$. The robustness of the
NPD- and GD-selected projected kernels is consistent with their shallow
circuits---a separable $Z$ feature map and a single-layer Hamiltonian
encoding---and their reliance on local observables, whereas the
Pareto-selected projected kernel uses a five-layer entangling IQP-style
encoding, whose depth makes it far more sensitive to noise.

Overall, these results show that the best-performing model in our study
is a fidelity quantum kernel, the NPD-selected Hamiltonian encoding (AUC
$0.840$), but its geometric difference ($g_k\lesssim 7$) remains well
below $\sqrt{N}\approx 14.14$, so a classical kernel cannot be ruled out
from matching its performance. Conversely, several projected quantum
kernels combine a large geometric difference with an AUC above the
classical baseline as seen in Fig.~\ref{fig:gd_auc}. None of them, however, reaches the
best-performing fidelity kernel, whose performance a classical kernel
may in principle recover. Performance and geometric separation are
therefore never achieved together at the top of the ranking: the
kernels that are clearly distinct from the classical reference are not
the best detectors, and the best detector is not clearly distinct. At
least for this dataset, we thus find no quantum kernel in our search
space that both certifiably departs from the classical reference and
outperforms every model a classical kernel could plausibly reach.

\subsection{Hardware results}
\label{sec:res-hw}
\begin{figure}
    \centering
    \includegraphics[width=0.8\linewidth]{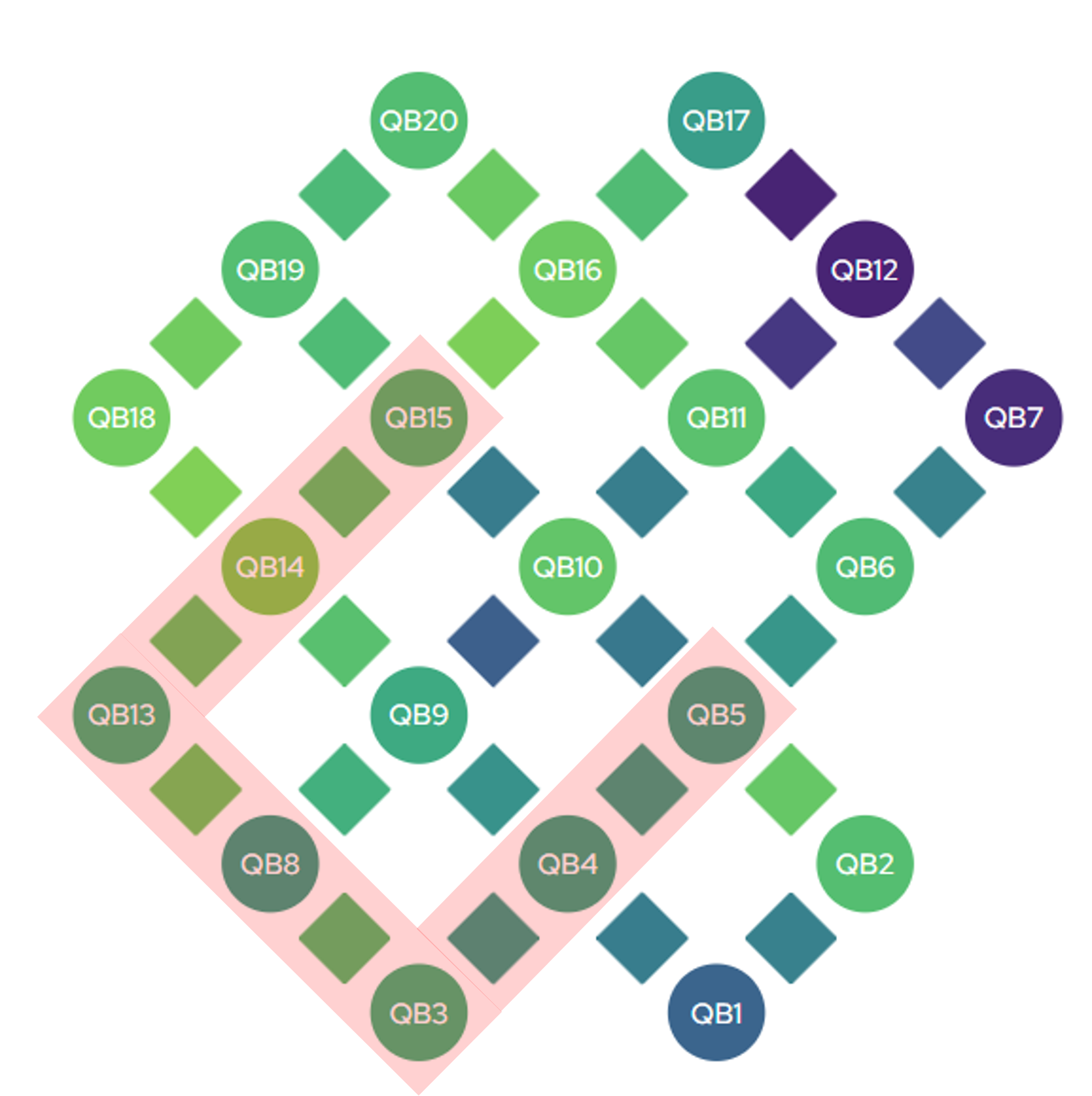}
    \caption{Layout of the IQM Garnet QPU. The color map indicates single and two-qubit gate errors (darker=higher error, lighter=lower error). We highlight the optimal way to choose a qubit patch to implement a Hamiltonian evolution feature map with nearest-neighbour interactions. The specific patch needs to be adjusted depending on re-calibration of the QPU.}
    \label{fig:layout}
\end{figure}

We complete our study with a benchmark of the best quantum models on IQM's QPUs to investigate the effect of noise on quantum kernel calculations and corresponding performance. We used IQM Garnet, a 20-qubit QPU based on superconducting transmon qubits. The qubits are arranged in a square lattice (see Fig.~\ref{fig:layout}) and connected by tunable couplers (more details about the coupling scheme can be found in \cite{marxer2023long}).
We performed two different experiments, using both fidelity and projected kernels and the optimal hyperparameters given by the optimization of the NPD score. In fact, this choice allowed us to have the best balance between quality of the results and practical implementation of the circuits on current hardware (i.e., shallower circuits). The results for the three performance metrics considered in this work are presented in Table~\ref{tab:exp_res}. The hardware results are computed on the $1{,}000$ test flows without per-pair aggregation, and are therefore not directly comparable to Table VI. For the projected kernel, $\nu$ was chosen as the value maximising the test scores, so its figures should be read as optimistic.\\
For fidelity kernels we implemented the Hamiltonian feature map with $2$ layers (i.e. $2$ Trotter steps - see App.~\ref{sec:supplementary} for more details) as reported in Table~\ref{tab:best_config_fidelity}. The optimal way to choose the qubits on IQM Garnet QPU is highlighted in Fig.~\ref{fig:layout}: the nearest-neighbour interactions of the Hamiltonian suggest the use of physically connected qubits in a 1D chain to optimize the depth of the transpiled circuits. Kernel entries in Eq.~\ref{eq:k} can be calculated with the \textit{adjoint} method (see e.g. \cite{hubregtsen2022training} and references therein), which doubles the depth of the circuit, but does not need auxiliary qubits. The number of circuits to run on the QPU to construct the training kernel matrix scales as $\mathcal{O}(N^2)$, where $N=200$ in our case; the number of shots per circuit was fixed to $1000$. To take into account measurement errors, we applied readout error mitigation to the probability distribution of bitstrings obtained with the QPU \cite{maciejewski2020mitigation, bravyi2021mitigating}. Finally, following \cite{hubregtsen2022training}, we further applied regularization techniques to ensure positivity of the kernel matrix and applied mitigation techniques tailored to depolarizing noise. Both the AUC score and the F1(N) decrease considerably when running the fidelity kernel on real hardware, while the F1(A) only differs slightly from the result in Table~\ref{tab:benchmark}. In this case, more advanced error mitigation and suppression techniques should be applied to improve the results, but this goes beyond the scope of the current work.\\ 
In the case of projected kernels, the data-encoding circuit is the Z separable map with $5$ layers (see Table~\ref{tab:best_config_pqk}). To calculate the single-qubit reduced density matrices in Eq.~\ref{eq:pqkdef} on the QPU, we used the classical shadows protocol introduced in \cite{huang2020predicting}. In this case, the number of circuits to evaluate scales linearly in the number of data. Nevertheless, we limited the size of the shadow to $100$, as the theoretical bounds in \cite{huang2020predicting} are not practically implementable on hardware (and are also expected to be much smaller in practice). The results reported in Table~\ref{tab:exp_res} show that the AUC metric is lower than the simulated value, while the F1 scores are higher. We believe that this behavior is due to the small size of the shadow rather than noise on the QPU, as the Z feature map only includes single-qubit gates and the circuits have low depth. To get higher precision for these results, one could use a bigger shadow size or apply a more recent protocol, multi-shots classical shadows, which has proven advantageous for Pauli measurements \cite{zhou2023performance} over the original method.\\
The results presented in this section leave room for future improvements, not only on the hardware run optimization side, but also on the search for optimal hyperparameters under more hardware-realistic noise models.

\begin{table}[h]
\centering
\caption{Performance of the quantum kernels obtained from the IQM Garnet QPU using the optimal configurations for the hyperparameters as given by the NPD score; for the projected kernel, $\nu$ was selected to maximise the test scores.}

\label{tab:exp_res}
\begin{tabular}{lcccc}
\toprule
Method & Feature Map & AUC & F1(A) & F1(N) \\
\midrule
Fidelity & Hamiltonian & 0.532 & \textbf{0.640} & 0.151 \\

Projected & Z & \textbf{0.612} & 0.630 & \textbf{0.656} \\

\bottomrule
\end{tabular}
\end{table}

\section{Discussion}
\label{sec:discussion}
The picture that emerges from our experiments is consistent but not the one a search for quantum advantage would hope for. The two objectives of our protocol are, on this dataset, in direct tension. Fidelity kernels sit at one end: NPD-guided selection finds a Hamiltonian-evolution feature map that improves on the tuned classical OCSVM baseline (AUC $0.840$ versus $0.765$), yet its geometric difference never exceeds $g_{C\rightarrow Q}\simeq 7$, below $\sqrt{N}=14.14$. By the criterion of Ref.~\cite{huang2021power}, this separation is too small to certify the improvement as quantum: the bound allows the reference classical kernel to close the gap with more training data (up to a factor $g^2\approx 49$ in the worst case), so the improvement carries no evidence of quantum origin.

Projected kernels sit at the other end, reaching $g_{C\rightarrow Q}$
up to $272$. Unlike in a purely negative
picture, several of them also detect better than the classical
baseline. However, none of them beats the
best fidelity kernel. The best-performing knee point of the Pareto optimisations is the
fidelity kernel (AUC $0.812$), which recovers most of the NPD-selected
performance ($0.840$) but stays in the low-GD regime. The projected
knee point reaches a lower AUC ($0.782$), though still the highest of
the three projected selections, at a GD well above $\sqrt{N}$; it
remains below the best fidelity kernel and is the projected
configuration most sensitive to noise (AUC $0.647$ on the IQM noisy
simulator).

This is the behaviour that the theoretical literature anticipates. A large geometric difference is necessary but not sufficient for a prediction advantage~\cite{huang2021power, kubler2021inductive}: it certifies that the quantum feature space is not well approximated by the chosen classical reference, but says nothing about whether the inductive bias it encodes is the one this task rewards. Our results are a concrete instance of that gap on real data, and they suggest that the missing ingredient is alignment between the feature map and the structure of network flows, not more expressivity. Trainable and data-dependent embeddings~\cite{hubregtsen2022training, glick2021covariant, altares2021automatic} offer a route to such alignment; in a fully unsupervised setting, the NPD could play the role that label-based kernel-target alignment plays in those works. The largest GD values in our study are also the least informative: they are reached by the projected kernel at the top of its bandwidth range, where $K_Q$ approaches the identity (Sec.~\ref{sec:res-fidelity}), so a large geometric difference can be produced by a single classical hyperparameter~\cite{shaydulin2022importance} and, on its own, is weak evidence of useful quantum structure.

Two aspects of the protocol itself are worth separating from these negative findings. First, NPD behaves as a usable label-free selection criterion for quantum kernels: the NPD-optimised configuration gave the best-performing model in two of the four settings (statevector fidelity and IQM projected) and came within $0.01$ AUC of the best in the other two (Table VI), which is the property one needs when labelled validation data is unavailable by construction. This should not be read as a monotonic relation between NPD and detection quality, within a family the two are only weakly related (Fig.~\ref{fig:npd_auc}), but as evidence that maximising NPD lands in a good region of the search space. Second, making GD a co-objective rather than a post-hoc diagnostic is what exposes the trade-off at all. A study reporting only NPD would have announced a quantum kernel beating the classical baseline; one reporting only GD would have announced a kernel outside the reach of the reference classical kernel. Both statements are true here, and neither on its own is informative.

The scope of these conclusions is limited in ways that matter. They rest on a single dataset with $N=200$ training flows and six engineered features, a regime dictated by the $\mathcal{O}(N^2)$ circuit cost of fidelity kernels on hardware; the $\sqrt{N}$ threshold is itself a function of this sample size, and a larger training set would raise the $\sqrt{N}$ threshold that the projected kernels currently exceed, although the geometric difference itself also depends on $N$. The Pareto fronts also reflect a finite budget of 100 trials per objective (150 for the fidelity kernel). The classical reference in the GD is the best kernel we found within a bounded family, so the separation is relative to that family and not to classical methods in general. Finally, the feature maps were taken from the literature rather than designed for network traffic, which is precisely the design freedom that our results identify as the binding constraint. In simulation the fidelity kernel outperforms the projected kernel (AUC $0.840$ versus $0.777$), but on IQM Garnet the situation is reversed (AUC $0.532$ versus $0.612$, on a different evaluation set, Sec. IV C): projected kernels degrade much less under hardware noise, consistent with shallower circuits and their reliance on local observables, although neither yet reaches the classical baseline ($0.765$).

We therefore read this work less as a negative result about quantum kernels on cybersecurity data than as a blueprint with an honest first application. The natural continuations are to construct feature maps informed by the periodicity and burstiness that characterise beaconing, rather than sampling generic encodings; to push $N$ upward using tensor-network or shadow-based kernel approximations~\cite{huang2020predicting} so that the $\sqrt{N}$ criterion becomes harder to satisfy trivially; and to apply the NPD--GD front to other unsupervised problems where labels are structurally unavailable. Whether any quantum kernel eventually occupies the upper-right corner of that front remains open; our contribution is a protocol that will make it visible when one does.


\section*{Acknowledgments}
This work was supported by the Région Île-de-France under the Pack Quantique programme (Grant No.~23002667). The authors gratefully acknowledge this financial support.

The authors thank the Allianz SE Cybersecurity and NextGen IT Think Tank, which led this collaboration, for domain expertise and critical review of the findings' relevance to industry security practice. The research was carried out by Multiverse Computing and IQM Quantum Computers using only publicly available data; no Allianz data, networks or production environments were examined. This exploratory work implies no control assessment, deployment, procurement or endorsement decision, and the views expressed are those of the authors and do not necessarily represent Allianz or its affiliates.

\bibliographystyle{apsrev4-2}
\bibliography{references}

\appendix
\section{Data-encoding feature maps}
\label{sec:supplementary}
We considered four different feature maps for encoding classical data into quantum states; in the following, the input data $x$ is a $n$-dimensional vector and $x_{j}$ is the $j$-th entry of $x$; $L$ denotes the number of layers.\\
\textbf{Z feature map} This feature map applies a wall of Hadamard gates and then parametrized Z-rotations to each qubit individually. It contains no entangling gates, thus the state at the end of the circuit is fully separable when applied to $\lvert 0^n \rangle$ state for $n$ qubits:
\begin{equation}
    \lvert \phi(x) \rangle
    =
    \left[
        \left( \bigotimes_{j=1}^{n} e^{- i Z_j \frac{x_{j}}{2} } \right) H^{\otimes n}
    \right]^{L}
    \lvert 0^n \rangle ,
    \label{eq:E1}
\end{equation}

\textbf{IQP-style feature map}  
This embedding was proposed in \cite{havlivcek2019supervised}, where the authors conjecture that the inner product resulting from this encoding is hard to compute classically, suggesting the possibility for a quantum advantage. The data-dependent state is defined as:
\begin{equation}
    \lvert \phi(x) \rangle
    =
    \left[ U_Z(x)\, H^{\otimes n} \right]^{L} \lvert 0^n \rangle ,
    \label{eq:E2}
\end{equation}
where $H^{\otimes n}$ applies Hadamard gates to all qubits in parallel, and
\begin{equation}
    U_Z(x)
    =
    \exp\!\left[
        -\frac{i}{2} \left(
        \sum_{j=1}^{n} x_{j} Z_j
        +
        \sum_{j<j'} x_{j} x_{j'} Z_j Z_{j'}
        \right)
    \right],
    \label{eq:E2_UZ}
\end{equation}

with $Z_j$ denoting the Pauli-$Z$ operator acting on the $j$-th qubit. The name IQP refers to \emph{instantaneous quantum polynomial-time} circuits \cite{shepherd2009temporally,bremner2011classical}: since $U_Z(x)$ is generated exclusively by Pauli-$Z$ terms, it is diagonal in the computational basis and all of its gates mutually commute. They can therefore be applied in any order, or formally as a single simultaneous (hence ``instantaneous'') interaction, so the data-dependent part of the circuit carries no temporal structure. The only non-commuting elements are the two data-independent Hadamard walls, which are what prevent the whole circuit from being classically simulable by sampling the diagonal generator directly.

\textbf{Hamiltonian evolution feature map}  
This feature map is inspired by quantum many-body dynamics and is implemented using a trotterization of the time evolution. It requires $n + 1$ qubits to embed $n$ features and is defined as follows (see \cite{huang2021power} for more details and references therein): 
\begin{align}
\lvert \phi(x) \rangle
&=
\Biggl[
\prod_{j=1}^{n}
\exp\!\left(
- i \frac{t}{T} x_{j}
\bigl(
X_j X_{j+1}
+ Y_j Y_{j+1}
+ Z_j Z_{j+1}
\bigr)
\right)
\nonumber
\\
&\qquad\quad
\Biggr]^{T}
\bigotimes_{j=1}^{n+1} \lvert \psi_j \rangle .
\label{eq:E3}
\end{align}
where $X_j$, $Y_j$, and $Z_j$ are the paulis acting on the $j$-th qubit, and $\lvert \psi_j \rangle$ are fixed Haar-random single-qubit quantum states. $t$ and $T$ are hyperparameters representing the total evolution time and the number of Trotter steps, respectively. 

\textbf{CNOT feature map} We also consider a hardware-inspired feature map based on single-qubit rotations and nearest-neighbour entanglement using controlled-$X$ (CNOT) gates. For a classical input vector $x$, the corresponding quantum state is prepared as
\begin{align}
\lvert \phi(x) \rangle
&=
\Biggl[
\prod_{\ell=1}^{L}
\left(
    \prod_{j=1}^{n} \mathrm{CNOT}_{j,j+1}
\right)
\left(
    \prod_{j=1}^{n} R_Z(x_j)
\right)
H^{\otimes n}
\nonumber
\\
&\qquad
\Biggr]
\lvert 0^n \rangle .
\label{eq:cnot_feature_map}
\end{align}
where $L$ denotes the number of layers, $H$ is the Hadamard gate, $R_Z(x_j)$ is a rotation around the $Z$ axis applied to the $j$-th qubit, and $\mathrm{CNOT}_{j,j+1}$ denotes a controlled-$X$ gate acting on neighboring qubits in a ladder-like topology.

\section{Feature engineering and feature selection}
\label{sec:features}
Network traffic was aggregated into unique flows defined by the tuple consisting of the \textit{source IP address}, \textit{destination IP address}, \textit{external destination port number}, and \textit{protocol}. For each unique flow, statistical features were then extracted. The objective here is to construct an effective set of features capable of capturing the complex patterns that separate benign network flows from malicious ones. The features generated for each flow are summarised below.

{\setlength{\parindent}{0pt}\setlength{\parskip}{3pt}
\textbf{\sffamily\small Flow duration.} Total flow duration (seconds).\par
\textbf{\sffamily\small Packet counts.} Total, forward, backward packets.\par
\textbf{\sffamily\small Packet length statistics.} Mean, std.\ dev., min, max, median, skewness, MAD (overall/fwd/bwd).\par
\textbf{\sffamily\small Flow rates.} Packets/s, bytes/s.\par
\textbf{\sffamily\small Inter-arrival time (IAT).} Min, max, mean, std.\ dev., total, coeff.\ of variation (overall/fwd/bwd).\par
\textbf{\sffamily\small Time-delta robustness.} Median, MAD, skewness.\par
\textbf{\sffamily\small Frequency-domain (FFT).} Mean, std.\ dev., max (overall/fwd/bwd).\par
\textbf{\sffamily\small Directional durations.} Forward and backward flow duration.\par
\textbf{\sffamily\small Directional rates.} Fwd/bwd packets/s and bytes/s.\par
\textbf{\sffamily\small Directional ratios.} Downstream/upstream (bwd-to-fwd) packet ratio.\par
}

Next, we need to select an optimal set of features from this list. This was achieved using hierarchical clustering~\cite{nielsen2016hierarchical}. The process can be described in two steps: 1) divide the large set of features into smaller groups of similar ones and 2) reduce the number of redundant features by selecting the most representative feature from each group.

We stress that this criterion is unsupervised: no label information enters the feature selection. We selected the optimal number of clusters for the hierarchical clustering using the silhouette score, which provides a quantitative measure of cluster quality by balancing intra-cluster cohesion and inter-cluster separation~\cite{rousseeuw1987silhouettes}. Specifically, we evaluated candidate numbers of clusters \(k\) by cutting the hierarchical dendrogram into \(k\) clusters and computing the silhouette score for each partition. The value of \(k\) that maximised the silhouette score was chosen as the optimal number of clusters, as it corresponds to the partition with the best overall separation structure. The motivation for this approach is that hierarchical clustering does not inherently determine the optimal cut level, and the silhouette score offers an objective, data-driven criterion to guide this choice.

From each cluster we then keep the medoid, i.e.\ the feature with the smallest average correlation distance to the other features in its cluster. This reduces the feature set to the $6$ features listed in Table~\ref{tab:features}.

\end{document}